\documentclass{wqcd03}                 
\usepackage{graphicx}
\confname{QCD@Work 2003 - International Workshop on QCD, Conversano, Italy,
14--18 June 2003}

\title{NON PERTURBATIVE EFFECTS IN QCD}

\author{Adriano Di Giacomo\addressmark{a}
       }

\address[a]{Dipartimento di Fisica Universita' Pisa and INFN Sezione di
Pisa, PISA  ITALY}

\begin{document}

\begin{abstract}
Non perturbative results from lattice QCD will be discussed, namely :
Vacuum Condensates and QCD Sum Rules; $U_A(1)$ and Topology; Confinement
of Color.
\end{abstract}

\maketitle


\section{INTRODUCTION}

QCD is usually quantized perturbatively.

The Lagrangean $L$ is split into
the sum of two terms
\begin{equation}
        L = L_0  +  L_I
\end{equation}
where $L_0$ describes free quarks and gluons ,and  $L_I$ their
interaction. Then $L_0$ is quantized : the Hilbert space is the Fock space
of free quarks and gluons and the interaction describes scattering
between them.

       Any observable $\langle O \rangle$ is  computed as a power series
expansion in
the renormalized coupling constant $\alpha_s$
\begin{equation}
       \langle O \rangle = \Sigma_n O_n \alpha_s^n
\end{equation}
where $O_n$ are finite amplitudes.

       The expansion eq(2) is not convergent, not even as an asymptotic
series\cite{H.Mueller}.
Nevertheless it is an empirical fact that, as $q^2 \to\infty$ or for $q^2
>      1Gev^2$
$\alpha_s\ll1$ and a few terms of the expansion describe physics
correctly. The lack of convergence of the expansion eq(2) reflects the
instability of Fock vacuum. Quarks and gluons are confined and therefore
free quarks and gluons are not a good zeroth order approximation to
physics.

A non perturbative quantization of the theory is needed to identify the
true vacuum state. This is provided by the Feynman path integral
formulation. The key quantity is the partition function $Z$, in terms of
which all field correlators can be computed.
\begin{equation}
Z = \int[dA_\mu][d\psi][d\bar\psi]\exp{-\left[S(A_\mu,\bar\psi,\psi)\right]}
\end{equation}
$Z$ in eq(4) is a functional integral and is defined by approximating
euclidean space time by a discrete set of points in a finite volume, so
that the number of integration variables is finite and the integral is an
ordinary well defined integral. A sequence of such integrals is then
computed on sets of points which tend to cover the volume densely. The
limit of infinite volume is then performed. If $Z$ is finite and well
defined after these limiting procedures one says that the theory exsists
as a field theory. The procedure also ensures the existence of the
analytic continuation to Minkowskian space time.

       For QCD asymptotic freedom insures that the first of these limits exists.
The existence of a mass gap in the theory also insures that the infinite
volume limit exists . A fully rigorous mathematical proof of these
statements still does not exist, but it is physically reasonable to say
that QCD exists as a field theory.

       Lattice formulation of  QCD is nothing but an approximant of Feynman
integral in the sequence which defines it : at sufficiently low value of
the unrenormalized coupling and at sufficiently large volume it will provide
the physical amplitudes with any required precision
from first principles (modulo technical computational difficulties).

Non perturbative phenomenology is based on Wilson's Operator Product
Expansion (OPE) ,

\begin{equation}
T(A(x) B(0))=\Sigma_nC_n(x) O_n(0)
\end{equation}
where $C_n$'s describe short distances and are usually computed in perturbation
theory and the operators $O_n$ describe large distances.The sum in eq(4)
is ordered by incresing order of the dimensions in mass of the local
operators $O_n$.

Vacuum expectation values (vev) of eq(4) involve in the rhs vev's of
composite operators (condensates) . They enter in the SVZ sum
rules\cite{SVZ}. The gluon condensate $\langle G_2\rangle=
\langle0|\beta(g_s)/g_s G^a_{\mu\nu}(0) G^a_{\mu\nu}(0)|0\rangle$ ,
and the quark condensate
$\langle G_{\bar\psi\psi}\rangle =
\langle0|\bar\psi(0)\psi(0)|0\rangle $ are examples of them.
A condensate like $G_2$ has dimension 4 in mass, and therefore, by
renormalization group arguments is related to the running coupling
constant at the scale $\mu$ as $G_2=\mu^4exp(-4/b_0/g^2(\mu^2))$
which is non analytic in g and undefined in perturbation theory.

Matrix elements of eq(4) between quark states are relevant to weak
interaction matrix elements at distances $\cong1/M_W,1/M_Z$ ,or to
structure functions of deep inelastic scattering.

In this talk I will concentrate on the vacuum state, namely on
condensates (sect 2) , topology (sect3) , and confinement of color (sect4)

\section{QCD SUM RULES. CONDENSATES}

The OPE of the product of two conserved currents

\begin{equation}
T(j_\mu(x)j_\nu(0) =  (  g_{\mu\nu} - \partial_\mu\partial_\nu)
\tilde{\Pi}(x^2)
\end{equation}
taken on the vacuum reads
\begin{equation}
\langle\tilde\Pi(x^2)\rangle\sim\tilde C_I^*(x^2) \langle I\rangle +
\tilde C_G(x^2)\langle G_2\rangle +
\tilde C_{\bar\psi\psi}(x^2)\langle G_{\bar\psi\psi}\rangle+..
\end{equation}
where I is the identity operator ( $\langle I\rangle=1$),and at short
distances $
C_I(x^2)\sim1/x^4$ modulo
logs,$ C_G$ and
$ C_\psi$
$\sim const$ modulo logs. Eq(6) is a theorem in perturbation theory, an
assumption in the presence of non perturbative effects. The coefficients
$\tilde C$ describe short distances, the condensates describe large distance
physics.

After Fourier transform
\begin{eqnarray}
&&\Pi(q^2) - \Pi(0)\equiv \int
d^4x\langle\tilde\Pi(x^2)\rangle(exp(iqx) - 1)\sim\nonumber\\
&&C_I(q^2) + C_{G_2}(q^2)\langle G_2\rangle +
C_{\bar\psi\psi}(q^2)\langle G_{\bar\psi\psi}\rangle
\end{eqnarray}
with
\begin{eqnarray}
&&C_I(q^2)\sim  const \qquad\hbox{(modulo logs)}\\
&&C_{G_2}(q^2),C_{\bar\psi\psi}(q^2) \sim 1/q^4 \qquad\hbox{(modulo logs)}
\end{eqnarray}
A dispersive representation for the l.h.s. of eq(7), if $j_\mu$ is the
electromagnetic current is
\begin{equation}
\Pi(q^2) - \Pi(0) = -q^2\int d\mu^2 \frac{R(\mu^2)}{\mu^2(q^2-\mu^2
+i\epsilon)}
\end{equation}
where $R(s)\equiv \frac {\sigma_{e^+e^-\to{hadrons}}}
{\sigma_{e^+e^-\to\mu^+\mu^-}}$ can be taken from
experiment.

An appropriate average on $q^2$ of both members of eq(10) gives the
SVZ\cite{SVZ} sum rules, in which
non perturbative effects are parametrized in terms of the condensates
$\langle G_2\rangle$ and $\langle  G_{\bar\psi\psi}\rangle$.
The result is
a good phenomenology and
a determination of the condensates
\begin{eqnarray}
&&\langle G_2\rangle = (.024\pm.011)Gev^4,\\
&&\langle G_{\bar\psi\psi}\rangle \sim -.13Gev^3     (q^2=1Gev^2)
\end{eqnarray}
However the perturbative expansion of $\Pi(q^2) - \Pi(0)$ which
provides the coefficient $ C_I(q^2)$,
when resummed at higher orders , is ambiguous by terms $\propto1/q^4$
\cite{H.Mueller},\cite{DoschV}
which mimic the terms with $G_2$ and $G_{\bar\psi\psi}$. The
definition of the condensates is then
intrinsically ambiguous.  They could be defined as the coefficients
of the terms $1/q^4$ in the
expansion eq(7), but a priori, due to the ambiguity, they could be
process dependent, i.e. dependent
on the currents under consideration.

An OPE of the levels of bound $\bar Q Q$ energy levels
\cite{leut}\cite{vol} can be also performed.
This amounts to compute the quadratic Stark effect produced on the
levels by the fluctuating vacuum
chromoelectric field. If the finite correlation length of the field
strength correlators is taken
into account\cite{gromes}\cite{CDO} the relevant quantity is the gauge
invariant correlator\cite{CDM}
\begin{equation}
\langle G_2(x)\rangle \equiv\langle 0|Tr\left\{\vec E(x)U_C(x,0)\vec
E(0)U^\dagger(x,0)\right\}|0\rangle
\end{equation}
with $U_C(x,0)$ the parallel transport from 0 to $x$ along the path C, or
$U_C(x,0)=
Pexp(i\int_CA_\mu(x)dx^{\mu})$.
       A general parametrization of such correlators is\cite{Do}\cite{SIM}
\begin{eqnarray}
&&\hskip-20pt D_{\mu\nu\rho\sigma}(x)=
\langle0|Tr[G_{\mu\nu}(x)U_C(x,0)G_{\rho\sigma}(0)U^\dagger_C(x,0)]|0\rangle=\nonumber\\
&&\hskip-20pt(g_{\mu\rho}g_{\nu\sigma} -
g_{\mu\sigma}g_{\nu\rho})[D(x^2)+D_1(x^2)] +\\
&&\hskip-20pt+\left[x_\mu x_\rho g_{\nu\sigma}-x_\mu x_\sigma
g_{\nu\rho}+x_\nu x_\sigma g_{\mu\rho}-x_\nu x_\rho
g_{\mu\sigma}\right]
\frac{\partial D_1(x^2)}{\partial x^2}\nonumber
\end{eqnarray}
       In this language $\langle G_2(x)\rangle=\alpha/\pi(D+D_1)$. The
correlators
$D_{\mu\nu\rho\sigma}$ can be
computed on the lattice \cite{CDM}\cite{DP} and with them their
OPE.
\begin{figure}
\includegraphics[scale=0.4]{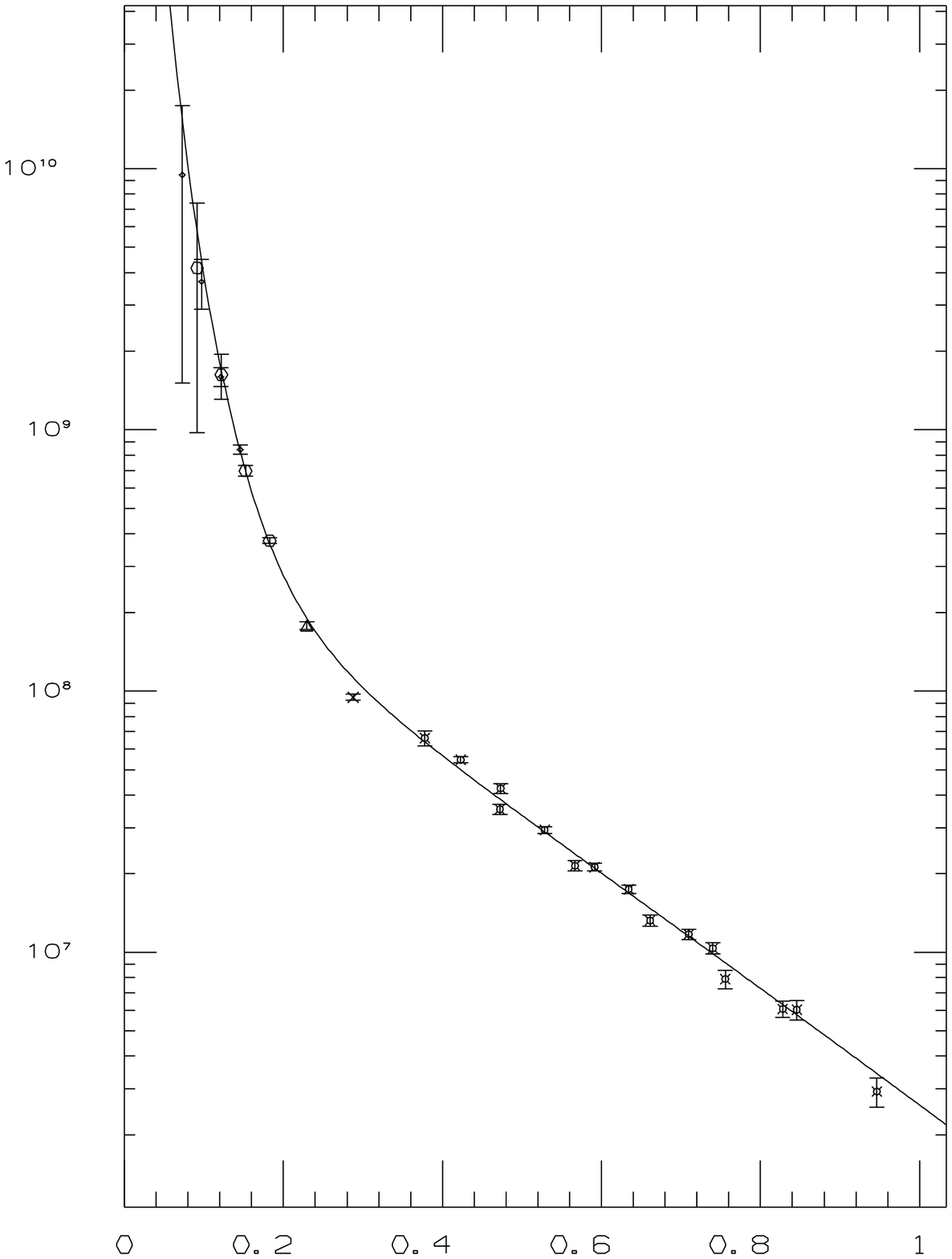}
\vskip-0.5in
\caption{$D^L(x)$. The line is the best fit to eq(15).}
\end{figure}
\begin{figure}
\includegraphics[scale=0.4]{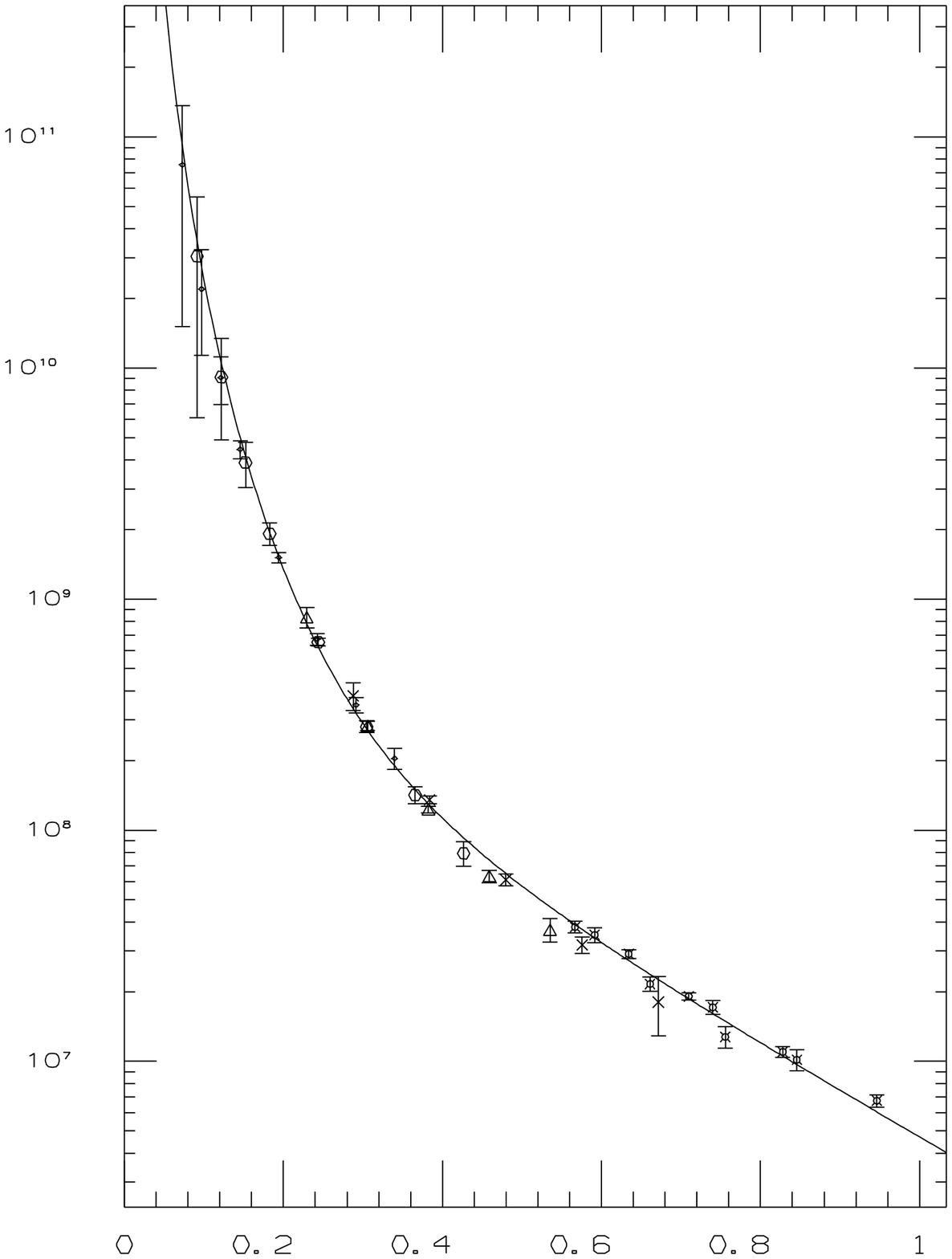}
\vskip-0.5in
\caption{$D^L(x)_1$. The line is the best fit to eq(16).}
\end{figure}
Their typical  behaviour is
shown in figs 1 and 2. A convenient parametrization of the Lattice data
$D^L, D^L_1$ is \cite{DP}\cite{DPM}
\begin{eqnarray}
&&\frac{1}{a^4}D^L
\mathop\simeq_{x\to0} \frac{C_I}{ x^4} + C_G\langle
G_2\rangle\exp(-x/\lambda)\\
&&\frac{1}{a^4}D^L_1  \mathop\simeq_{x\to0} \frac{{C'}_I}{ x^4}
+ {C'}_G \langle G_2\rangle\exp(-x/\lambda')
\end{eqnarray}
$\langle G_2(x)\rangle$ is a split point regulator of $\langle G_2\rangle$.
      A best square fit to the lattice data gives, for pure gauge
SU(3)\cite{DPM} (quenched,no quarks)
$G_2=(.15\pm.03) Gev^4, \lambda=\lambda'=(.22\pm.03)fm$ .

In full QCD \cite{DDM} the same quantities can be computed. An
extrapolation to realistic quark
masses gives $\langle G_2\rangle=(.022\pm.005)Gev^4$ and
$\lambda=\lambda'=(.32\pm.04)fm$.
The phenomenological value for $\langle G_2\rangle$ is
$(.024\pm.005Gev)^4$\cite{nar}

     From quenched to full QCD the gluon condensate decreases by almost an
order of magnitude,and the
correlation length increases by 50\%.

Lattice vacuum correlators are the input of the Stochastic approach
to QCD \cite{SIM}\cite{adgDSS}.

In a similar way the fermion correlators can be computed\cite{DEDGM}
\begin{equation}
       \langle G_{\bar\psi\psi}(x)\rangle\equiv \langle
0|\bar\psi(x)U_C(x,0)\psi(0)|0\rangle
\end{equation}
$ \langle G_{\bar\psi\psi}(x)\rangle$ can be viewed as a split point
regulator of
$\langle G_{\bar\psi\psi}\rangle$.

       Lattice data \cite{DEDGM} give $\langle0|\bar\psi\psi|0\rangle$
consistent with the value
obtained by use of SVZ sum rules
and a correlation length $\lambda=(.42\pm.05)fm$

\section{TOPOLOGY}
The $U_A(1)$ problem was a problem of the free quark model of
Gellmann. The singlet axial current is
conserved in the model, but in Nature the corresponding symmetry is
neither realized a la Wigner
(no parity doublets exist in the hadron spectrum), nor a la Goldstone
: indeed a Goldstone-broken
symmetry would imply  \cite{W} $m_{\eta'} < \sqrt{3} m_{\pi}$ ,which is
badly violated by the observed
value $m_{\eta'}\sim980 Mev$. Hence the idea of Gellmann that the
symmetry of hadrons could be
abstracted from the free quark model was not correct for the axial singlet.

QCD solves the puzzle. The singlet axial current is anomalous in QCD
and $U_A(1)$ is not a
symmetry,being broken at the quantum level.

\begin{equation}
\partial_\mu j^{\mu}(x) =2N_fQ(x)
\end{equation}
      $Q(x) = \frac{1}{32\pi^2}G_{\mu\nu}G^*_{\mu\nu}$

($G^*_{\mu\nu}\equiv 1/2\epsilon_{\mu\nu\rho\sigma} G_{\rho\sigma}$
) is named  topological charge
density.

On continuous configurations  the topological charge $Q= \int d^4x
Q(x)$ is an integer (second
Chern Number).

\begin{figure}
\includegraphics[angle=270,scale=0.3]{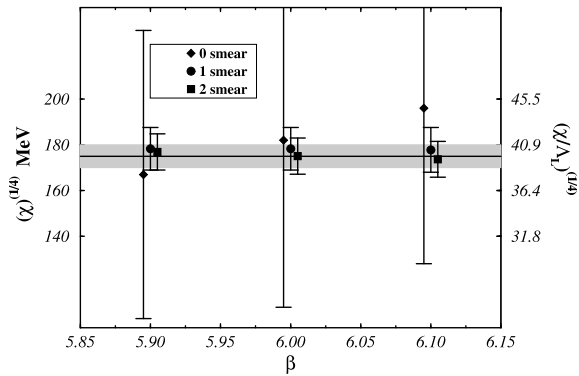}
\caption{Determinations of $\chi$ by use of different lattice regulators
and of different values of the unrenormalized coupling constant. The
physical value of $\chi$ is independent of these choices within
statistical errors. }
\end{figure}

For quantized fields $Q(x)$ is an operator.
      The Topological susceptibility $\chi$ is defined as the response of
the vacuum to Q(x)
\begin{equation}
\chi= \int d^4x <0|T(Q(x)Q(0))|0>
\end{equation}
An argument based on $N_c\to\infty$ shows that $m_{\eta'}$ is related
to the topological
susceptibility of the $N_c=\infty$ vacuum.\cite{witten}
\begin{equation}
2N_f\chi=f_\pi^2 m_{\eta'}^2 ( 1+ O(1/N_c))
\end{equation}
      As $N_c\to\infty$ dynamical quarks
can be neglected and with the same approximation $N_c$ can be taken
equal to the physical value 3
,so that $\chi$ is the topological susceptibility of the quenched SU(3) vacuum.
      An improved version of the argument including mixing to the octet
gives\cite{ven}
\begin{equation}
2N_f\chi = f_{\pi}^2 (m_{\eta'}^2+m_{\eta}^2-2m_K^2) (1+O(1/N_c))
\end{equation}
The  prediction is in both cases $\chi\sim  (180Mev)^4$.

Computing $\chi$ from first principles can provide a cross check of
QCD and of the 1/$N_c$ expansion
at the same time.

      This problem has a long story: tha main difficulty has been for a
long time to understand how
topology, which is based on continuity, can exist on the lattice
which has a discrete structure.
      The way out is basic field theory\cite{CDP} : any lattice version of
the operator $Q(x)$, $Q_L(x)$,
is related to the continuum counterpart ,say in the $\overline{MS}$ scheme,
by a multiplicative
renormalization
\begin{equation}
Q_L(x) = Z_QQ(x)
\end{equation}
The product Q(x)Q(0) is singular as $x\to0$.The OPE gives
\begin{eqnarray}
&&Q(x)Q(0)\sim  c_I(x) I + c_2(x) G_2 + \hbox{finite terms}\\
&&c_I\sim x^{-8} , c_2\sim x^{-4}
\end{eqnarray}
      In computing $\chi$ an additive renormalization is also
required.\cite{DFRV}
In conclusion\cite{pippo}
\begin{equation}
\chi=   \frac{\chi_L-\chi_L^0}{Z^2}
\end{equation}
where $\chi_L^0$ is the lattice topological susceptibility of the
$Q=0$ sector of the vacuum.
A technique developed to compute $Z_Q$ and $\chi_L^0$ is known as
"heating"\cite{DGV}: it is based
on the fact that topology has a much longer autocorrelation time
$\tau_Q$ in montecarlo upgrading
than the autocorrelation time $\tau$ of the local quantum
fluctuations which dominate the
renormalization constants. Therefore $\chi^0_L$ can be measured on the
configurations obtained at
times t such that $\tau\ll t\ll\tau_Q$ from a zero field
configuration, which has topological charge
zero. Similarily $Q_L$ can be measured on configurations prepared
from an initial configuration
consisting of a single instanton , which has Q=1. At times long
enough with respect to $\tau$ but
very short compared to $\tau_Q$ $Q_L$ will be equal to  Z . This
method is known as
"field-theoretical".
      The result of this procedure is  $\chi=((175\pm5)Mev)^4$\cite{ADD}
Recently an independent determination has been made, using the
divergence of the singlet axial
current operator,i.e. the lhs of eq (17). The advantage is in the use
of a formulation of the
fermions on the lattice which preserves chiral symmetry, and hence
has multiplicative
renormalization $Z_Q=1$ The result is\cite{GW}
\begin{equation}
\chi=((190\pm10)Mev)^4
\end{equation}
in complete agreement with the field theoretical determination.
Also determinations based on cooling techniques \cite{teper} give
consistent estimates.
In conclusion the $U_A(1)$ problem is solved in QCD, and additional
legitimation is added to the
ideas of $N_c\to\infty$.
      An important issue, as we shall see in the next section, is the
behaviour of $\chi$ at the
deconfining phase transition.

\section{CONFINEMENT OF COLOR}
Quarks and gluons are visible at short distances, but have never been
observed as free particles.
This fact has lead to the conjecture that colored particles never
appear in asymptotic states,
a property known as Confinement of Color. Confinement should be
derivable from the QCD lagrangean.

Quarks have been searched in Nature since they were introduced by
Gellman as fundamental
constituents of hadrons\cite{GEL}, with negative results: only upper
limits to their abundance and
to their production rate have been established\cite{PDG}.

The ratio of quark abundance in nature, $n_q$ to that of protons
$n_p$ has been given the upper limit
\begin{equation}
\frac{n_q}{n_p}\le10^{-27}
\end{equation}
     to be compared to the expectation based on the Standard Cosmological
Model  $\frac{n_q}{n_p}\approx10^{-12}$ .

As for production in particle reactions the best limit is provided by
the inclusive cross section
$\sigma_q\equiv\sigma(p+p \to q(\bar q) + X )$  namely
\begin{equation}
\sigma_q\le 10^{-42}cm^2
\end{equation}
to be compared with the expectation, in the absence of confinement,
\begin{equation}
\sigma_q\simeq\sigma_{TOTAL}\approx10^{-25}cm^2
\end{equation}
In both cases the ratio is depressed by a factor $\sim10^{-15}$. The
only natural explanation is
that these ratios are exactly zero, i.e. that confinement is an
absolute property, and as such based
on some symmetry.

Experiments have been and are being performed to detect a deconfining
transition at high
energy density from hadronic matter to a state of free quarks and
gluons (Quark Gluon Plasma),
in high energy collisions of heavy ions\cite{QM}. The major
difficulty with them is to identify an
observable quantity by which deconfinement could be detected.

Some evidence for the existence of that transition has been produced
by virtual experiments, namely
by numerical simulations of QCD on a lattice.

The partition function of a field theory at temperature T is the
euclidean Feynman path integral
extending in euclidean time from $\tau=0$ to $\tau=\frac{1}{T}$,
with periodic boundary conditions
for boson fields, antiperiodic for fermions.

For QCD this is computed
on a lattice of spatial
extension $N_s^3$ and temporal extension $N_t$ with $N_t\ll N_s$, and
the temperature is
$T=\frac{1}{a(\beta)N_t}$ with $a(\beta)$the lattice spacing in
physical units , which can be tuned
by varying the unrenormalized coupling constant
$\beta\equiv\frac{2N_c}{g_s^2}$.

Also on the lattice, however, as in experiments, the real problem is
to have a criterion to detect
confinement.

In quenched formulation (pure gauge, no dynamical quarks) the
criterion consists in looking at the
large distance behaviour of the static $\bar QQ$ potential, $V(\vec
x)$ . $V(\vec x)$ is related to
the correlator of Polyakov lines $D(\vec x)$
\begin{equation}
     D(\vec x) = \langle L(\vec x) L^\dagger(0)\rangle
\end{equation}

    by the relation
\begin{equation}
V(\vec x) = -\frac{1}{aN_t}ln(D(\vec x))
\end{equation}
where $aN_t$ is the extension of the lattice in the time direction.
     By definition  the Polyakov line is the trace of the parallel
transport across the lattice in the
time direction.
\begin{equation}
L(\vec x) = P exp[i\int _{0}^{1/T}d\tau A^0(\vec x,\tau)]
\end{equation}
By cluster property
\begin{equation}
D(\vec x)
\mathop\simeq_{x\to\infty}
exp(-\sigma x/T) + |\langle L \rangle|^2
\end{equation}
If $\langle L \rangle=0$ it follows
\begin{equation}
V(\vec x)\sim \sigma x   \quad\hbox{as\quad}  x\to\infty
\quad\hbox{(confinement)}
\end{equation}
If $\langle L \rangle\neq0$
\begin{equation}
V(\vec x)\sim const.    \quad\hbox{as\quad}  x\to\infty
\quad\hbox{(deconfinement)}
\end{equation}
A transition is observed on the lattice at $T_c\approx 270Mev$ from a
region $T < T_c$ where $\langle L \rangle$=0 to a region $T > T_c$ where
$\langle L \rangle\neq0$. A finite size scaling analysis allows to
determine the order of the transition, which for SU(3) is
first\cite{COL}\cite{TS}.
$\langle L \rangle$ can be then assumed as an order parameter for
confinement , and the corresponding symmetry is $Z_3$

In principle one should show that $\langle L \rangle =0$ implies absence
of colored particles in all asymptotic states to prove confinement, which
has not been done, but the criterion is reasonable anyhow.

In the presence of dynamical quarks (full QCD) $Z_3$ is not a symmetry
anymore and  $\langle L \rangle$ is not an order parameter. String
breaking is expected to take place: when pulling the static $\bar QQ$
pair apart from each other energy is converted into light $\bar qq$ pairs,
and the potential is not linear any more, even if there is confinement.

At $m_q=0$ another symmetry is present, chiral symmetry, which is
spontaneously broken at $T=0$ , and, as lattice data show, is restored at
some $T_c$. Chiral symmetry, however, is explicitely broken by quark
masses, and cannot be the symmetry responsible for confinement.

Let us consider the case $N_f=2$, $m_u=m_d=m$, which is semirealistic,
but,as we shall see, very instructive.

A phase dyagram is usually drawn
for this system as in fig(4). The line which joins the first order phase
transition at $m=\infty$ (quenched) to the chiral transition  at $m=0$
corresponds to the values $T_c(m)$ at which the susceptibilities
$C_V$ (specific heat) and $\chi_{\bar \psi\psi} =\int d^3x\langle \bar
\psi(\vec x)\psi(\vec x)\bar \psi(0)\psi(0) \rangle$ are
maximum\cite{KL}\cite{TS2}.

\begin{figure}
\includegraphics[scale=0.3]{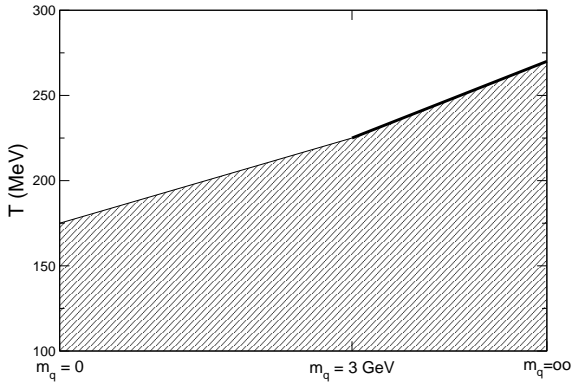}
\caption{Schematic phase diagram of $N_f=2$ QCD. The line corresponds to
maxima of the susceptibilities. }
\end{figure}

    Conventionally the region below this line
is called confined, the region above it deconfined. No criterion exists
which justifies this attribution.

An analysis can be made of the chiral phase transition at $m=0$, assuming
that the pions and sigmas are the relevant degrees of freedom\cite{PW}.
Chiral symmetry fixes then the form of the free energy and
renormalization group arguments allow to predict the order of the
transition and the corresponding universality class.

For $N_f=3$ the transition is first order. For $N_F=2$ the order
depends on the relative weight  of the  chiral O(4) part of the action and
the term describing the $U_A(1)$ anomaly. If the anomaly disappears below
$T_c$ the transition is first order, and such is the transition along the
line at $m\ne 0$. If instead the anomaly persists up to $T_c$ the
transition is second order the universality class is $O(4)$ and the line
at $m\ne 0$ is a crossover. A reliable numerical analysis of this issue is
not yet available, but for some reason the second option is more popular.

     The order and the universality class can be investigated by studying the
behaviour of the critical specific heat as the volume goes to infinity
by a finite size scaling analysis. A good order parameter should have
a behaviour in that limit consistent with that of $C_V$ .The order of the
transition and the universality class should be identified by the
symmetry responsible for confinement.

     A valid candidate symmetry is dual superconductivity of the
vacuum\cite{tH75}\cite{MA}\cite{tH81}. The basic idea is that the
chromoelectric field acting between colored particles is channeled into
dual Abrikosov flux tubes, whose energy is proportional to the length.
Here dual means interchange of electric with magnetic with respect to
ordinary superconductors.

This mechanism has been assumed as a working hypothesis and analyzed
down to observable consequences in a series of papers
\cite{ZAK}\cite{DDP}\cite{DDPP}\cite{DLMP}\cite{DLMP1}\cite{DP}\cite{DPA}.

A disorder parameter $\langle \mu \rangle$ has been defined which detects
dual superconductivity: it is the vev of an operator  $\mu$ carrying
magnetic charge which has zero vev in a phase in which vacuum has a
definite magnetic charge, and can have a non zero vev if monopoles
condense i.e. the vacuum is a superposition of states with different
magnetic charge.

    That operator is color gauge invariant , magnetically
charged but magnetic U(1) gauge invariant\cite{DP}.

In principle $\langle
\mu \rangle\ne 0$ or$\langle \mu \rangle = 0$ i.e. superconductivity
or not depends on the procedure used to identify magnetic charges (Abelian
Projection). However it can be shown that monopole condensation (or non
condensation) is an abelian projection independent
statement\cite{D}\cite{DPA}.

For quenched theory the result of numericall simulatios is that this
disorder parameter is consistent with $\langle L
\rangle$\cite{DLMP}\cite{DLMP1}  : it is non zero for $T < T_c$ and
strictly zero for $T > T_c$. By finite size scaling analysis of the
corresponding susceptibility $\rho=\frac {d}{d\beta} log(\langle \mu
\rangle)$ the same critical indices are obtained as with $\langle
L \rangle$.

The operator $\mu$ is perfectly well defined also in the
presence of dynamical quarks, and can be used to investigate the phase
diagram of fig(4). The result is that the region below the line is really
confined,the one above it is deconfined\cite{ref3}.

$\rho$ as a
function of $T$ at fixed $m$ has a peak at the same value as the other
susceptibilities, i.e. on the line of fig(4).
     $\langle \mu \rangle$ thus provides a good criterion for confinement.

The finite size scaling analysis works as follows. A priori $\langle \mu
\rangle$ depends on the coupling constant $\beta$ i.e. on the lattice
spacing $a$ , on the mass $m$   and on the
lattice size $N_s$  .The dependence on
$\beta$ can be traded with the dependence on the correlation length $\xi$
so that, for dimensional reasons
\begin{equation}
\langle \mu \rangle = \Phi(\frac{a}{\xi},\frac{N_s}{\xi},mN_s^{\gamma})
\end{equation}
As $T$ approaches $T_c$ from below, or $\tau\equiv (1-\frac{T}{T_c})\to0$,
$\xi$ diverges as
\begin{equation}
\xi\propto \tau^{-\nu}
\end{equation}
    $\nu$ and $\gamma$ are critical indices .

If the transition is
second order or weak first order $\xi$ goes large as $T_c$ is approached
and
$\frac{a}{\xi}$ can be put equal to zero. Because of eq(37)
$\frac{N_s}{\xi}$ can be traded with $\tau N_s^{1/\nu}$ . The scaling law
follows
\begin{equation}
\rho/N_s^{1/\nu} = f(\tau N_s^{1/\nu},m N_S^{\gamma})
\end{equation}
In the quenched case the second dependence does not exist.The scaling
can be tested on the lattice data and the critical index $\nu$ can be
extracted\cite{DLMP}\cite{DLMP1}. In particular for the peak one finds
\begin{equation}
\rho_{peak}\propto N_s^{1/\nu}
\end{equation}

For the case of full QCD the problem has two scales and the analysis is
more complicated. Preliminary results\cite{PICA} exclude the critical
indices of
$O(4)$ and seem consistent with a first order phase transition.
If this will be confirmed by the additional data in preparation
either the anomaly disappears below $T_c$ , or the relevant degrees of
freedom which dominate are not the pion and the sigma\cite{PW}.

     The first possibility can be checked on the lattice. The dependence of
the topological susceptibility at the deconfining transition has been
studied\cite{ADD}.Fig(5) shows the existing data. A more precise
analysis is needed to settle the problem.

\begin{figure}
\includegraphics[angle=270,scale=0.4]{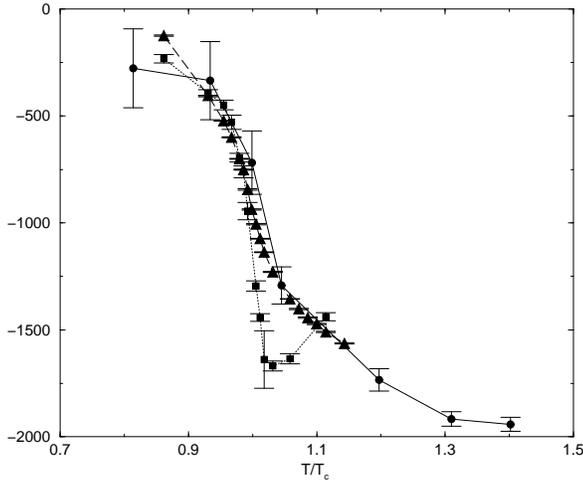}
\caption{The behaviour of the topological susceptibility (cercles), of
the magnetic monopole susceptibility $\rho$ (squares), and of the chiral
parameter $\langle \bar\psi\psi \rangle$ (triangles) versus $T/T_c$ at
the deconfining transition. }
\end{figure}

     In conclusion there is strong evidence that dual superconductivity is
the mechanism of confinement. A candidate disorder parameter exists ,
which is well defined and seems to work. The corresponding symmetry is
dual superconductivity,which is however
not clearly understood formally, being a  common feature of all the
abelian projections: consequently the effective action is not known.

A definite settlement of the order of the transition for the case
$N_f=2$ will be a crucial test.

\end{document}